\documentclass[submission,copyright,creativecommons]{eptcs}
\providecommand{\event}{ACL2 2022} 
\usepackage{breakurl}              

\title{VWSIM: A Circuit Simulator}
\author{Warren A. Hunt, Jr., Vivek Ramanathan, and J Strother Moore \\
\institute{The University of Texas at Austin}
\institute{Computer Science Department and ForrestHunt, Inc. \\
Austin, Texas \ \ USA}
\email{\{hunt,vivek,moore\}@forresthunt.com}
}

\def\titlerunning{VWSIM: A Circuit Simulator}
\def\authorrunning{Hunt, Ramanathan, and Moore}

\usepackage{threeparttable}

\usepackage{pgfplots} 

\usepackage{graphicx}        

\usepackage{amsmath,amssymb,amsfonts,amsthm}
\usepackage{tikz}

\usetikzlibrary{calc, trees, positioning, arrows, arrows.meta, automata,
  chains, decorations.markings, decorations.pathreplacing,
  decorations.pathmorphing, matrix, shapes, shapes.gates.logic.IEC,
  shapes.gates.logic.US, shapes.geometric, shapes.symbols}

\tikzset{ basic/.style = {draw, text width=2cm, font=\sffamily,
    rectangle}, level 2/.style = {basic, rounded corners=6pt, thin,
    align=center, fill=green!60, text width=8em}, level 3/.style =
  {basic, rounded corners=6pt, thin, align=center, fill=orange!60,
    text width=8em}, level 4/.style =
  {basic, rounded corners=6pt, thin, align=center, fill=yellow!60,
    text width=12em}, level 5/.style =
  {basic, rounded corners=6pt, thin, align=center, fill=yellow!60,
    text width=8em} }

\tikzset{>={Stealth}}

\tikzset{->-/.style={decoration={
  markings,
  mark=at position .5 with {\arrow{Stealth}}}, postaction={decorate}}}

\tikzstyle{branch}=[fill, shape=circle, minimum size=4pt, inner sep=0pt]

\tikzstyle{clk} = [regular polygon, regular polygon sides=3,
  inner sep=1pt, text centered, draw=black]

\tikzstyle{cond} = [diamond, minimum width=0.8cm, minimum
  height=0.8cm, text centered, draw=black]

\tikzstyle{data} = [rectangle, minimum width=1cm, minimum height=0.7cm,
  text centered, draw=black, fill=yellow]

\tikzstyle{l} = [draw, -latex']

\tikzstyle{mux} = [trapezium, draw, shape border rotate = 270,
  trapezium angle = 60, inner ysep=0pt, outer sep=1pt, inner xsep=1pt,
  text width = 1.5em]

\tikzstyle{operation} = [circle, minimum width=0.8cm, minimum
  height=0.8cm, text centered, draw=black]

\tikzstyle{process} = [ellipse, minimum width=1.4cm, minimum height=1cm,
  text centered, draw=black]

\tikzstyle{reg} = [rectangle, minimum width=1.5cm, minimum height=0.7cm,
  text centered, draw=black, fill=orange!60]

\usepackage{xcolor}
\tikzset{>=latex} 
\colorlet{Icol}{blue!50!black}
\colorlet{Rcol}{violet!90!black}
\colorlet{Ccol}{blue!90!black}
\colorlet{Bcol}{black!90!black}
\colorlet{pluscol}{red!60!black}
\colorlet{minuscol}{blue!60!black}
\tikzstyle{thick C}=[C,thick,color=Ccol,Ccol,l=$C$]
\tikzstyle{thick R}=[R,thick,color=Rcol,Rcol,l=$R$]
\tikzstyle{thick W}=[-*,thick,color=Bcol,Bcol,l=$R$]

\usepackage[american]{circuitikz}

\begin{document}
\maketitle

\begin{abstract}

VWSIM is a circuit simulator for rapid, single-flux, quantum (RSFQ)
circuits. The simulator is designed to model and simulate
primitive-circuit devices such as capacitors, inductors, Josephson
Junctions, and can be extended to simulate other circuit families,
such as CMOS. Circuit models can be provided in the native VWSIM
netlist format or as SPICE-compatible netlists, which are flattened
and transformed into symbolic equations that can be manipulated and
simulated. Written in the ACL2 logic, VWSIM provides logical
guarantees about each of the circuit models it simulates. Note, our
matrix solving and evaluation routines use Common Lisp floating-point
numbers, and work is ongoing to admit these models into ACL2. We
currently use VWSIM to help us design self-timed, RSFQ-based
circuits. Our eventual goal is to prove properties of RSFQ circuit
models. The ACL2-based definition of the VWSIM simulator offers a path
for specifying and verifying RSFQ circuit models.

\end{abstract}



\section{Introduction}   





We have defined the VWSIM circuit simulator with simulation models for
resistors, capacitors, inductors, transmission lines, mutual
inductance, Josephson Junctions (JJs), and VWSIM includes voltage,
current, and phase sources.  VWSIM can simulate an entire circuit
model either in the voltage or phase domain.  With our use of
ACL2~\cite{ACL2} to implement the VWSIM electrical-circuit simulator,
VWSIM enjoys mathematical properties not available in other simulation
systems: termination, logical consistency, and a formal semantics for
reasoning about circuit models.

The VWSIM system was authored using the ACL2 formal logic; thus the
entire simulator implementation can be thought of as a mathematical
function that animates circuit models.  Given suitable input stimulus,
VWSIM can produce voltage, current, and phase records.  Developing
tools using ACL2 provides a number of benefits.  Each ACL2 function,
when defined, must be proven to terminate and proven to access
reachable data only; thus, all ACL2 programs are known to terminate,
without memory-reference errors.

From an ACL2 perspective, the main contribution of this work is the
definition of a tool for rapid, single-flux, quantum (RSFQ) circuit
development; the definition and guard verification of a SPICE-like
simulator; and the push it has provided to include IEEE-compatible,
floating-point numbers and associated functions within the ACL2 core.
Our presentation will touch on these points.

The VWSIM system accepts SPICE-compatible~\cite{SPICE} circuit models
so that it can read existing circuit descriptions.  VWSIM also accepts
input in its native Lisp format; this format provides ACL2-specific
mechanisms for specifying input circuit models, stimulus, and
requirements that can be checked during simulation.  VWSIM provides
various primitives for specifying stimulus input; this language allows
a user to create input waveforms by referencing existing generators or
by user-provided stimulus functions.

When our effort began, we identified properties we hoped RSFQ modeling
and analysis would provide.  We realized that available simulators for
RSFQ circuits were all void of some properties we desired.  These
shortcoming did not impede our use of public-available simulators, but
no simulator combined all of the properties we felt were important for
investigating RSFQ circuit models thoroughly and accurately.  Note, we
have and continue to use JoSIM~\cite{JoSIM, Delport-2019}; the JoSIM
simulator provides us with inspiration, and we strive to make our
system as good as JoSIM in areas where JoSIM shines while providing
capabilities not available in any available circuit simulator.  Some
of our goals for VWSIM include:

\begin{enumerate}
\item
{\bf ACCURATE:} We wrote VWSIM with clarity and simplicity in mind.
\item
{\bf RELIABLE:} VWSIM is proven to terminate without memory-access errors.
\item
{\bf FLEXIBLE:} VWSIM offers a general-purpose stimulus language.
\item
{\bf INTEGRAL:} VWSIM supports voltage and phase-based
simulation.\footnote{Phase is a concept widely used in superconducting
systems; phase is proportional to the time integral of voltage.  Phase
is a potential.  The current through an inductor is proportional to
the phase across the inductor divided by its inductance.  The current
through a JJ is proportional to the sine function of the phase across
it.  Phase is specified in units of radians.}
\item
{\bf TIMING:} Fixed and variable time-step simulations are available.
\item
{\bf ACCESSIBLE:} All code is freely available; runs on multiple systems.
\item
{\bf INTEROPERABLE:} Use of VWSIM can be scripted with ACL2 or other tools.
\item
{\bf CAN BE PAUSED:} Any simulation may be paused, saved, and later re-started.
\item
{\bf EXTENSIBLE:} VWSIM can be extended with additional components and stimulus.
\item
{\bf ANNOTATION:} VWSIM models can be annotated with physical parameters.
\item
{\bf PROMPT:} VWSIM can be used interactively; signal traces are memory resident.
\item
{\bf FORMAL:} The VWSIM is written in a mathematical modeling language.
\item
{\bf INTROSPECTIVE:} VWSIM source code has been analyzed with the ACL2 system.
\item
{\bf ANALYSIS:} VWSIM code and circuit models can be analyzed symbolically by ACL2.
\end{enumerate}

VWSIM is written in a manner that allows a user to inspect and confirm
the veracity of its internals.  When a circuit is analyzed by VWSIM, a
set of equations is produced through an analysis of the input model;
these equations are produced using the Modified Nodal Analysis (MNA)
method~\cite{MNA-1975}.  This abstract model is instantiated and
evaluated each simulation cycle so that circuit models are sensitive
to simulation node and branch values, time-varying external stimulus,
and even time-varying component values.

An unusual feature of VWSIM is that its definition has been analyzed
by the ACL2 theorem proving system.  During the development of VWSIM,
hundreds of guard theorems were proven to assure that VWSIM operates
reliably; our guard specifications and proofs help assure that our
data structures remain well-formed.  This ability was an important
component of our development process as it allowed us to confirm
properties of functions used in VWSIM operation.  In addition, it is
even possible to analyze the behavior of VWSIM with respect to
user-supplied model input; we have not explored the capability with
respect to VWSIM models, but we have performed such work
~\cite{FM9001,Micro-Verification} on other projects.



Below is a very simple, RC circuit schematic.  This circuit is
composed of three elements; a voltage source, a resistor, and a
capacitor.  Lines (wires) in a schematic are given names, such as {\tt
  vs1}, {\tt vc1}, and {\tt gnd} as shown in our upcoming example.
But every actual wire in a superconducting (SC) circuit behaves and
stores energy like an inductor.  Thus, we have to distinguish
equipotential nodes and inductive wires.

\begin{figure}
  \begin{center}
    \begin{tikzpicture}

      \draw (0,2) to[short,*-, l=$vs1$] (0.1,2.0)
      to[short,-]             (0.9,2.0)
      to[thick R,  l=$R_1$] ++(1.5,0)
      to[short,-          ] ++(0.9,0)
      to[short,-* ,l=$vc1$] ++(0.2,0)
      to[short,- , l=     ] ++(0.9,0)
      to[thick C,  l=$C_1$] ++(1.5,0)
      to (6,2)
      -- (6,2) -- (6,0)
      to[short,-,] ++(-5.8,0)
      to[short,-*,l=$gnd$] (0.2,0);

      \node at (-0.2,1) {$\Delta \hspace{0.2cm} v1$};
      \node[below left] at (0,2) {$+$};
      \node[above left] at (0,0) {$-$};
    \end{tikzpicture}
  \end{center}
  \caption{An RC circuit with a 1-Ohm resistor, $R_{1}$, and a 1-Farad capacitor, $C_{1}$. }
\end{figure}
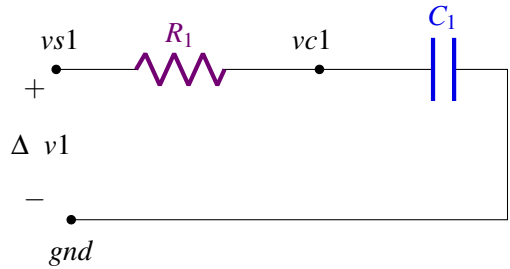


We represent the schematic diagram above in our native netlist format
as shown below.  The name of this netlist is {\tt *rc-netlist*} and it
contains one module, {\tt rc-module}.  This module contains no external
connections, as indicated by the {\tt nil} in its second line.  This
module contains three components: a voltage source, a 1-ohm resistor,
and a 1-farad capacitor.  The first field in the component statements
are occurrence names; they must be disjoint.  The second field
specifies the type of components; in this netlist there is a voltage
source ({\tt v}), a resistor ({\tt r}), and a capacitor ({\tt c}).
The third field specifies the nodes to which each component is
connected. The fourth field contains names to reference the current
that passes through the individual components.  The final (fifth)
field specifies the value(s) for each component.

\begin{verbatim}
(defconst *rc-netlist*
  '((rc-module
     nil
    ; Name type  connections branch   value
     ((v1    v    (vs1 gnd)  (i-v1)   ((if ($time$< '1/5) '0 '1)))
      (r1    r    (vs1 vc1)  (i-r1)   ('1))
      (c1    c    (vc1 gnd)  (i-c1)   ('1))))))
\end{verbatim}

The voltage source, {\tt v1}, is specified to provide zero volts until
the time advances to 1/5 second.  The voltage source {\tt v1} has an
annotation that describes its specified behavior: while simulation
time is less than one 1/5 second, the voltage it produces is zero;
once the time reaches one 1/5 of a second, the voltage source produces
a one-volt output for the rest of the simulation.

As shown below, one may check whether a netlist is well-formed using
the {\tt NETLIST-SYNTAX-OKP} and {\tt NETLIST-ARITY-OKP} predicates.
The {\tt vwsim} form specifies a simulation starting at time 0,
proceeding with a 1/5-second simulation time step until two seconds of
simulation has been completed using a {\tt voltage}-style simulation.
The results of the RC-circuit simulation are just below.  This
capacitor-charge simulation shows voltage {\tt vc1} approaching one
volt.



\begin{verbatim}
(netlist-syntax-okp *rc-netlist*)
(netlist-arity-okp  *rc-netlist*)

(vwsim *rc-netlist*
       :time-step 1/5
       :time-stop 2
       :sim-type voltage)
\end{verbatim}

\begin{samepage}
\begin{verbatim}
(($TIME$ 0.00 0.20 0.40 0.60 0.80 1.00 1.20 1.40 1.60 1.80)
 ($HN$ 0.00 0.20 0.20 0.20 0.20 0.20 0.20 0.20 0.20 0.20)
 (I-V1 0.00 -0.91 -0.74 -0.61 -0.50 -0.41 -0.33 -0.27 -0.22 -0.18)
 (I-C1 0.00 0.91 0.74 0.61 0.50 0.41 0.33 0.27 0.22 0.18)
 (GND 0.00 0.00 0.00 0.00 0.00 0.00 0.00 0.00 0.00 0.00)
 (VS1 0.00 1.00 1.00 1.00 1.00 1.00 1.00 1.00 1.00 1.00)
 (VC1 0.00 0.09 0.26 0.39 0.50 0.59 0.67 0.73 0.78 0.82))
\end{verbatim}
\end{samepage}

VWSIM uses floating-point numbers, but ACL2 provides only rational
arithmetic natively.  We note that there is an ongoing effort to
modify ACL2 so it can use floating-point numbers to approximate ACL2's
native rational-number arithmetic by appealing to contemporary,
(IEEE-compatible~\cite{IEEE_754}) floating-point implementations; we
discuss this more in Section~\ref{sec:future-work}.

The rest of this paper provides our usage model and outlines the
process of converting a list of components, along with their
interconnections, into a model that we can simulate by repeatedly
solving for $x$ in the matrix equation $Ax = b$.  We describe how we
create a symbolic form of the $Ax = b$ equation that we instantiate
repeatedly solving for a new value of $x$ each time step.  We describe
how the simulation values are stored so that further processing can be
performed after, or even during, simulation.  In fact, analyzing the
state of the simulator during its operation allows VWSIM to modify
simulation input as a function of previous simulation states.  We
conclude with a comparison of VWSIM to JoSIM, which served as an
inspiration of our work.

\section{Background}


Circuit simulation has a long history.  SPICE~\cite{SPICE} was defined
in the 1970s to provide circuit simulation; SPICE and is derivatives
have been used broadly in circuit simulation for fifty years.  Other
simulators, such as WRspice~\cite{WRspice} and PSCAN2~\cite{PSCAN2}
are also used for superconducting circuit simulation.  Certainly,
there are many other simulators, but we do not survey them here.

We began the development of our simulator for several reasons: to
improve our understanding of the mathematics of JJ-based circuits, our
failure to understand what existing simulators were doing, a lack of a
way to program existing simulators to carry out collections of
simulations, the lack of a way to alter a simulation after it is
started, the lack of an interactive means to control the simulation
process, the lack of a programmed method for inspecting results, and
the lack of a formal semantics for JJ-based circuits.  As we have
developed our simulator, we have learned that specifying the
interconnections of components does not provide an adequate model for
the behavior of in JJ circuits.  For instance, transformers are
modeled by coupling two inductors, and JJs have a non-linear behavior,
where tracking the phases, as well as the voltages and currents, is
critical for performing accurate simulations.

The design of our simulator was strongly influenced by
JoSIM~\cite{Delport-2019} -- a simulator designed for simulating
circuits containing Josephson Junctions.\footnote{We had several very
helpful communications with Joey Delport, the JoSIM author.}
Initially, we conducted our study of superconducting circuits using
JoSIM as a simulation engine for circuit model we created using the
Electric CAD system~\cite{GNU-Electric}.  The development of our
simulator has progressed to the point where we use it in all instances
except when simulating large circuits because JoSIM executes several
times faster that VWSIM.

During our use of JoSIM, we encountered several JoSIM simulation
results we could not explain nor understand -- and, in most of these
instances, we eventually submitted a bug report.  In one instance, we
were misled for nearly a month before we were able to distill our
confusion down to a model so small that we could compare our by-hand
simulation with JoSIM results.  Eventually, we decided that the best
way we could really understand how an electric circuit simulator for
JJs operates was to develop our own simulator.  We did not make this
decision lightly.  The development of VWSIM has consumed more than a
person-year of effort.



\section{How VWSIM Operates} \label{sec:the_process}


VWSIM accepts a hierarchical circuit description as a list of circuit
modules, which it then flattens into a list of primitive devices with
no module (subcircuit) references. The flat netlist is checked for
consistency, making sure that there are no undefined references or
unrecognized components.  With a valid, flat netlist, VWSIM attempts
to build a matrix equation that represents a symbolic set of
relations; these relations include components (e.g., a specific
resistor) and sources (e.g., some time-varying current source), where
their specific values are to be instantiated during simulation.  If
the resulting matrix, $A$, is well-formed (i.e., not singular) and the
sources, in $b$, are suitable, then VWSIM prepares to solve for $x$ in
matrix equation $Ax = b$ by instantiating the symbolic expressions in
$A$ and $b$ with the current simulation values.  The solution vector,
$x$, represents the next-time values for the simulation variables for
a time-step value specified by the user -- this process is repeated
until simulation time is exhausted.  VWSIM records the voltages and/or
phases of all wires (nodes) as well as the currents through all
components except resistors and JJs.  The currents through resistors
and JJs can be calculated after the simulation, if requested. A VWSIM
user can then process the simulation history, either by inspection, by
analysis with an ACL2 program, or by visualizing some projection of
the recorded information.  Later, we describe this process in more
detail.

\begin{figure}
\resizebox{15.5cm}{6.5cm}{%
\begin{tikzpicture}[
    node distance=2cm,
    inp/.style = {align=center, font=\linespread{1}\selectfont, line width=0.3mm},
      alg/.style = {draw, align=center, font=\linespread{1}\selectfont, line width=0.3mm},
      io/.style={draw, trapezium, trapezium left angle=70, trapezium right angle=110, minimum width=1cm, minimum height=1cm, align=center, line width=0.3mm},
      decision/.style={draw, aspect=2, diamond, minimum width=2cm, minimum height=0.5cm, align=center,line width=0.3mm},
    ]
    \node (node0) [alg]                             {VWSIM netlist \\ SPICE .cir file \\ .lisp file};
    \node (node1) [alg, right of=node0,xshift=1.5cm]  {Parse Input \\ into \\ VWSIM Modules};
    \node (node2) [alg, right of=node1,xshift=1.6cm]  {Sort and \\ Flatten Modules};
    \node (node3) [alg, right of=node2,xshift=0.9cm]  {Create \\ Symbolic \\ Ax=b};
    \node (node4) [io, below of=node3,yshift=0.1cm]  {Simulation \\ Loop};
    \node (node5) [alg, below of=node4, yshift=0.3cm]   {Increment time \\ with time-step amount};
    \node (node6) [decision, below of=node5, yshift=-0.2cm] {time $>=$ end time?};
    \node (node7) [alg, right of=node6, xshift=2.4cm]   {Evaluate Symbolic \\ A and b};
    \node (node8) [alg, right of=node7, xshift=1.2cm]   {Solve Ax=b};
    \node (node9) [alg, below of=node8, yshift=0.7cm]   {Record new \\ simulation values};
    \node (node10) [alg, below of=node6, yshift=0cm]   {Process and save results};

    \draw [arrows=-Stealth] (node0) --node[anchor=north]   {} (node1);
    \draw [arrows=-Stealth] (node1) --node[anchor=east]    {} (node2);
    \draw [arrows=-Stealth] (node2) --node[anchor=east]    {} (node3);
    \draw [arrows=-Stealth] (node3) --node[anchor=west]    {} (node4);
    \draw [arrows=-Stealth] (node4) --node[anchor=west]    {} (node5);
    \draw [arrows=-Stealth] (node5) --node[anchor=north]   {} (node6);
    \draw [arrows=-Stealth] (node6) --node[anchor=north]    {No} (node7);
    \draw [arrows=-Stealth] (node7) --node[anchor=west]    {} (node8);
    \draw [arrows=-Stealth] (node8) --node[anchor=east]    {} (node9);
    \draw [arrows=-Stealth] (node9) |- ([shift={(3mm,-8mm)}]node9.east)
    |- (node5.east);
    \draw [arrows=-Stealth] (node6) --node[anchor=west]    {Yes} (node10);
\end{tikzpicture}
}
\caption{VWSIM flowchart}
\label{sec:process-flowchart}
\end{figure}
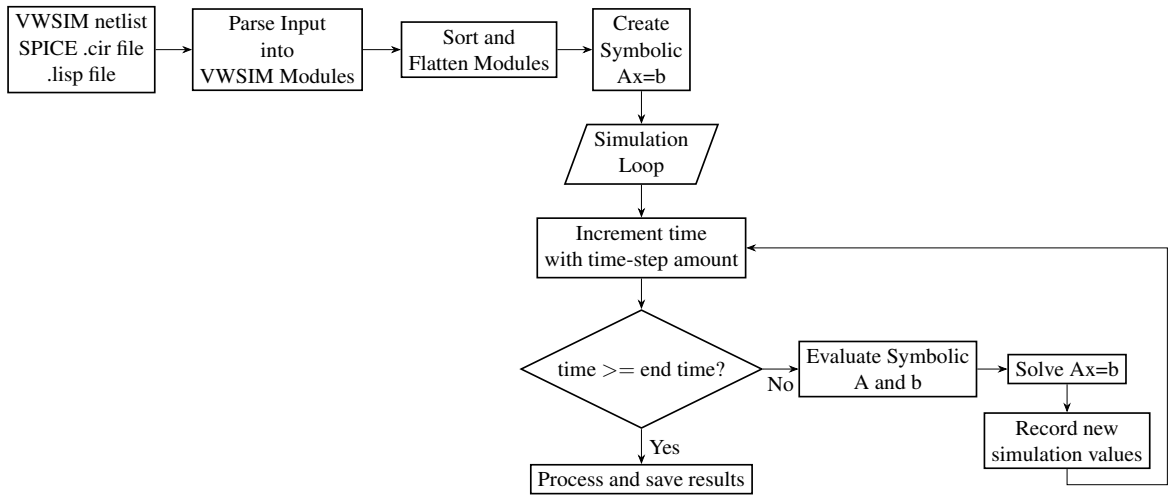

Given a textual description of a circuit, VWSIM will process it (see
Figure~\ref{sec:process-flowchart}):
\begin{enumerate}
\item
  Convert input file, if in SPICE-format, into S-expressions.

\item
  Transform S-expressions into list of circuit module references and
  simulation control statements.

\item
  Convert module references into a hierarchical netlist representation.

\item
  Flatten and sort hierarchical netlist into a list of primitive
  device references.

\item
  Create a symbolic version of the model; i.e., create the symbolic
  $Ax = b$ matrix using the modified nodal approach
  (MNA)~\cite{MNA-1975} procedure.

\item
  Given a time step size (or using an automatically selected time step
  size), solve for $x$ in $Ax = b$.  Use these new $x$ values to
  extend the simulation, and repeat this process until the end of
  simulation time.

\item
  Finally, process and/or save the time-value results produced by the
  simulation; for instance, print results into a file of circuit-node
  values or check the simulation results for specific relationships.
\end{enumerate}

Our effort involved the initial development of a simple, but slow
simulator that we used to make sure that our approach was sound.  We
represented arrays as lists of lists.  We used rational numbers for
time and simulation state variables.  We kept the relations in a
symbolic form so we could read the equations produced by the MNA
method.  This approach allowed us to understand the operation of this
kind of simulator thoroughly, but the execution performance was
abysmal.  The MNA method produces a matrix that is related to the size
of the circuit, where two or three equations are needed for each
component; transformers and delay lines require additional equations.
Solving for $x$ in $Ax = b$ is cubic in the dimension of $x$.  After
simulating circuits with a few tens of components, we were waiting for
results that never appeared.  This is not surprising considering we
were using linear lists to represent our arrays; we started this way
because it was simple.

We then began an incremental replacement of our list-based data
structures for $A$ with a sparse matrix.  This change also caused a
complete update of our solving code so it could operate correctly on
data represented in a sparse matrix.  In turn, each of these changes
required increasingly difficult guard proofs.  In fact, these guard
proofs continue as we attempt to prove the guards of a sparse matrix
solving code that is coded with ACL2's second-order {\tt LOOP\$ FOR}
iteration constructs.  We further discuss our sparse-matrix solver
later in our paper.
\section{Using the VWSIM Simulator}     





This section explains how to run the simulator, presents the data
structures used to compute and store the simulation state, and
demonstrates the programmatic capability of VWSIM.


\subsection{Running the simulator}

The simulator is invoked by running the {\tt vwsim} function.  VWSIM
requires the user to provide a SPICE (.cir) circuit description file,
VWSIM-style circuit netlist or a LISP file with a previous VWSIM
simulation state.  The rest of the arguments are optional with default
values. The optional arguments enable the user to perform a voltage-
or phase-mode simulation, produce symbolic equations ($Ax = b$) that
can be inspected, save or resume a simulation, and filter and present
simulation output.  The function signature and formal parameters are
detailed below.
\begin{verbatim}
(vwsim <input>
       :sim-type <sim-type>
       :equations <equations>
       :spice-print <spice-print>
       :global-nodes <global-nodes>
       :time-step <time-step>
       :time-stop <time-stop>
       :time-start <time-start>
       :output-file <output-file>
       :concat-char <concat-char>
       :save-sim <save-sim>
       :save-sim-shortp <save-sim-shortp>
       :load-sim <load-sim>
       :save-var <save-var>
       :return-records <return-records>
       )

\end{verbatim}

where

\begin{itemize}
\item
{\tt input}: a Lisp-style VWSIM netlist or string with a filename.

\item
{\tt sim-type}: simulation mode; select {\tt voltage} or {\tt phase}.

\item
{\tt equations}: if non-NIL, return symbolic simulation equations
instead of a numerical simulation.

\item
{\tt spice-print}: whether the output file should contain only the
values requested in the SPICE file.

\item
{\tt global-nodes}: list of globally-defined nodes (wires) in the
circuit.

\item
{\tt time-step}: simulation time step in seconds as a rational number.

\item
{\tt time-stop}: simulation stop time in seconds as a rational number.

\item
{\tt time-start}: simulation start time in seconds as a rational number.

\item
{\tt output-file}: if provided, the filename to write the requested
SPICE simulation values.

\item
{\tt concat-char}: SPICE hierarchical circuit concatenation
character for PRINT statements.

\item
{\tt save-sim}: if provided, the filename to save the simulation state
for resuming later.

\item
{\tt save-sim-shortp}: if provided, save/write single-precision simulation values.

\item
{\tt load-sim}: if non-NIL, {\tt input} is a ``.lisp'' file
providing a previous state and simulation is resumed.

\item
{\tt save-var}: {\tt ASSIGN} variable name to save simulation results.

\item
{\tt return-records}: return specific (voltage, current, phase) simulation records.
\end{itemize}

\subsection{VWSIM simulation state}

VWSIM calculates, manipulates, and stores the simulation state using
five STOBJs (single-threaded objects): {\tt ABR}, {\tt ST-VALS}, {\tt
  DZ}, {\tt RTIME}, and {\tt REC}.

\begin{itemize}
\item The {\tt ABR} STOBJ contains 31 fields, which store most of the
  simulation state.  The first few fields are the symbolic A and b
  matrices that are evaluated and used to solve for the unknown
  voltages, currents, and/or phases of the circuit. The rest of the
  fields contain simulator configuration information.

\item The {\tt ST-VALS} STOBJ contains a single array that stores the
  results of evaluating every symbolic expression in the sparse A
  matrix and in the b vector stored in the ABR STOBJ.

\item The {\tt DZ} STOBJ is used by the linear system solver, which
  stores a sparse representation of the numerical A matrix and the b
  vector.

\item The {\tt RTIME} STOBJ contains two lists: the exact (rational)
  simulation times and the (rational) simulation time steps.

\item The {\tt REC} STOBJ contains a single array where all simulation
  variables and their values are stored for each time step.
\end{itemize}

When the first four steps of the process (section
\ref{sec:the_process}) are completed, the function {\tt simulate}
completes the initialization of the fields in {\tt ABR}, {\tt
  ST-VALS}, {\tt DZ}, and {\tt REC} fields and then begins the
transient analysis of the circuit provided.
\begin{verbatim}
    (simulate Abr st-vals dz rtime rec)
\end{verbatim}
The {\tt simulate} function proceeds by first instantiating the
symbolic equation representing the circuit model; this is discussed in
the next subsection.  After the $Ax = b$ equation has been
instantiated, VWSIM then attempts to solve this equation numerically,
see subsection~\ref{sec:ax=b_solver}.

\subsection{Symbolic Terms and Their Evaluation}


As discussed earlier, VWSIM maintains a symbolic version of the
equations of each simulation it is asked to process.  On each
simulation step, VWSIM instantiates each variable with present
simulation values so as to calculate the next simulation state.  Each
symbolic expression is evaluated given bindings for the equation
variables.

The {\tt vw-eval} function is an evaluator for the symbolic
expressions that are created when the input model is analyzed.  {\tt
  vw-eval} provides semantics for a number of real-valued functions,
such as addition ({\tt f+}), sine ({\tt f-sin}), and exponentiation
({\tt f-exp}).  ACL2 avoids offering definitions for $sin(x)$ where
$x$ is a real valued number.  Our primitive definitions of real-valued
functions appeal to Common Lisp's implementations for these functions.

\subsection{The $Ax = b$ Solver}\label{sec:ax=b_solver}

We wrote a series of linear equation solvers to solve $Ax = b$.  The
reason we wrote a series of them was to get a solver that was fast
enough for sparse matrices that are, say, 2000 x 2000. We are, of
course, aware that there are many extraordinarily fast linear equation
solvers.  We wrote one in ACL2 because ultimately we wish to prove it
correct.  We also experimented with several ACL2(p) implementations to
take advantage of parallelism but have so far had more success with
sequential implementations.

Our current solver is a Gaussian elimination solver.  There are many
variations on the basic Gaussian elimination method, but we started
with a particularly simple one as follows.  First, augment each row of
A with the corresponding entry from b to represent an entire equation.
Second, use the first row to cancel out the leading coefficient of
each subsequent row by scalar multiplication and addition.  Then recur
on the remaining rows, all of which have a leading zero.  Provided a
solution exists, this reduces the set of equations to triangular form
without changing the solution.  Then use back substitution to compute
the solution, starting at the bottom-most row and working up.

Our current solver is just an optimization of this basic method.  It
uses a STOBJ to store $A$, $x$, and $b$ as arrays.  An important
optimization when dealing with floating-point data is to permute the
rows so as to select as the ``pivot'' row one whose leading non-zero
coefficient has the largest absolute value.

We do not augment $A$ with $b$ but instead put $A$ into triangular
form and then compute from it a program that says how to solve for $x$
given any $b$.  The program, stored as an ACL2 list in the STOBJ, is
just a list of instructions recording the row indices and coefficients
used in the operations of the back substitution.  This allows us to do
most of the work just once and then use the triangular form repeatedly
for the different $b$'s generated by the simulator.

Since our matrices are sparse, we mirror A with a matrix, Z, that
records how many zeros are between the current non-zero value of A and
the next non-zero value in the current row.  This allows us to skip
over typically long gaps when canceling.



We are currently verifying the guards of this ACL2 solver.  The most
difficult part has been proving that the graph encoded in evolving $A$
matrix remains acyclic as we permute its rows.  Knowing that $A$ is
cycle free is necessary to prove the termination of most array update
functions.

\subsection{Optimizing the simulator}

The first version of VWSIM was very slow. In fact, it was barely able
to manage small circuit descriptions.  In order to handle increasingly
larger circuit descriptions, several optimizations were implemented to
reduce the time and memory consumption of a simulation.

\textbf{Sparse Matrix Representation of Symbolic Equations}: when
circuit descriptions are converted to the equation $Ax = b$, the $A$
matrix is usually very sparse.  In order to leverage this property,
the simulator only stores the non-zero entries in each row of the $A$
matrix. The $A$ matrix is an array, where each entry (row) is an
association list of pairs with a column index and a value.  For
example, {\tt '((0 . 1) (5 . 2))} is row with a value of {\tt 1} in
column {\tt 0}, {\tt 2} in column {\tt 5}, and {\tt 0} in all other
columns.  This matrix storage format reduces the amount of memory
required drastically, while also decreasing the access and update
times for each entry in the $A$ matrix.

\textbf{Using STOBJs to store the simulation state}: Single-threaded
objects, STOBJs, enable VWSIM to persistently store the simulation
state, which allows for saving and resuming simulations, filtering and
analyzing results, and more.  Additionally, STOBJs provide support for
arrays in ACL2.  VWSIM makes extensive use of arrays to enjoy
constant-time access and storage.  The $A$ matrix, $b$ vector, the
simulation output, and symbolic term evaluation results are all stored
in arrays.  This makes accessing and updating the simulation state
faster than linked-list-style structures.

\textbf{Symbolic subterm evaluation}: to perform a simulation, all
unique symbolic entries (terms) in the $A$ and $b$ matrices are
gathered.  From these symbolic terms, a list of all of their unique
subterms is generated.  For example, the unique subterms of the term
{\tt '(f* (f- '4 y) (f- '4 y))} are
{\tt
 Y,
 '4,
 (F- '4 Y), and
 (F* (F- '4 Y) (F- '4 Y))}.
The list of all subterms are ordered so that each term's subterms are
before it. This enables efficient evaluation of all unique terms in
one sweep, and no term is evaluated more than once.  The results are
stored in the array field of the {\tt ST-VALS} STOBJ.  Hence, when the
symbolic A and b matrices are being evaluated, each entry's evaluated
value is quickly accessed from the array.

\textbf{Fast, sparse-matrix solver}: The first version of VWSIM used a
rational-arithmetic-based solver that did not leverage the sparsity of
circuit simulation matrices.  This led to our effort to develop a
faster solver that would be capable of handling larger $A$ matrix
sizes.  This work has already been mentioned previously in Section
\ref{sec:ax=b_solver}.

\textbf{Floating-point approximation}: The native ACL2 rational
arithmetic provides arbitrary precision at the cost of performance.
As the lengths of simulations increase, the number of operations
performed on the rational values increases.  Since the precision of
rational numbers is unbounded, the numerators and denominators
continue growing, which increases memory usage and decreases
performance.  Additionally, VWSIM makes use of several approximation
techniques (trapezoidal integration method), irrational numbers
($\pi$, $e$), and transcendental functions ($\sin(x)$, $e^{x}$), for
which the use of arbitrary precision is superfluous.  Thus, VWSIM
makes use of double-precision, floating-point arithmetic to exploit
hardware-support for fast floating-point operations.

\subsection{Floating-point Simulations}

A VWSIM simulation uses and produces floating-point numbers.  Due to
the lack of support for floating-point arithmetic in ACL2, we employ a
\textit{trick} to ensure that the simulator can still be defined in
ACL2.  We define a recognizer, {\tt nump}.  In the ACL2 logic, {\tt
  nump} is equivalent to {\tt rationalp}; however, in raw lisp, a
double-precision floating-point number is recognized as a valid {\tt
  nump}.

\begin{verbatim}
    (defun nump (x)
      (declare (xargs :guard t))
      (and (acl2-numberp x)
           (zerop (imagpart x))))

    (defthm nump-is-rationalp
      (equal (nump x) (rationalp x)))
\end{verbatim}

VWSIM uses exact, rational numbers for the time and the time-step
values.  We note that these time values are artifacts of our
simulator; real circuits do not have time values and VWSIM does not
permit irrational time values.  VWSIM takes special care to use and
store the simulation time and time-step values precisely as rational
numbers.  This ensures that any time-sensitive events, such as input
sources that provide stimulus at specific times, occur exactly when
intended.

\section{An Example Circuit Simulation and Analysis}   



For RSFQ circuits, the active element is the Josephson Junction (JJ).
Energy for JJ operation is provided by a bias current; typically, JJs
are biased to about 70\% of the current required to cause their
unique, non-linear {\em firing} operation.  We present a basic
Josephson-style D-latch, discuss its operation, and show how we
represent and simulate this circuit when using VWSIM.

\begin{figure}




\begin{center}
\begin{circuitikz}[american,
                   scale=0.8
                   ]
  \ctikzset{bipoles/thickness=1}

  \draw [thick] (0,0) node[anchor=north]{D}
                      to [barrier, o- ] (2.5,0) 
                (1.25,0.15) node[above] {$J_3$};  

  \draw [thick] (2,1.8) to [short, i>=$I_{bias}$] (2,0);

  \draw [thick] (2,0) to [barrier] (2,-2)
                (2.1,-1) node[right] {$J_1$};  

  \draw [thick] (2,-2) node[tground]{};

  \draw [thick] (2,0) to [L, *-*, l={L}] (5,0);
  \draw [-, thick] (2.8,-0.15) to (4.2,-0.15); 

  \draw [thick] (5,2) node[anchor=east]{C}
                      to [barrier, o- ] (5,0)     
                (5.1,1) node[right] {$J_4$};  

  \draw [thick] (5,0) to[barrier] (5,-2)
                (5.1,-1) node[right] {$J_2$};  

  \draw [thick] (5,-2) node[tground]{};

  \draw [thick] (5,0) to (6,0);
  \draw [thick] (6,0) to [L, -o, i>=$ $] (9,0) node[anchor=north]{Out};

\end{circuitikz}
\end{center}  

\caption{The D flip-flop is a simple RSFQ ``logic'' element.  The
  input JJ, $J_3$, rejects input pulses when data (current) is
  circulating in the $J_1$ --L-- $J_2$ loop; otherwise, such an input
  causes $J_1$ to fire, which then causes some current to be captured.
  When a current is circulating, a pulse on input $C$ will cause $J_2$
  to fire, creating a current on terminal $OUT$.}
\label{fig:d-flip-flop}
\end{figure}
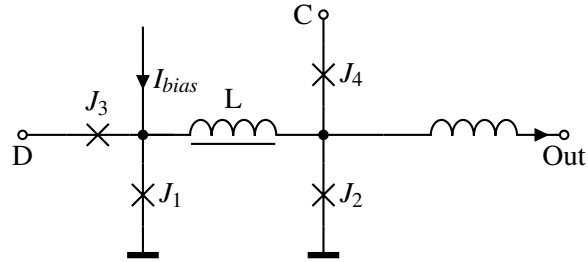

A basic RSFQ circuit element is the D-latch, which is the state-holding
device\cite{Mukhanov:1987} shown in Fig.~\ref{fig:d-flip-flop}.  Our
example is a two-junction interferometer with two stable states: one
with little-to-no current circulating and one with current circulating
in the $J_1$ -- $L$ -- $J_2$ loop.  Because of the superconducting
nature of these circuits, such a current will circulate indefinitely.
The circulating current is caused by the insertion of a certain amount
of flux into the loop \cite{bunyk:2001}.
The magnitude of the fluxon is quantized, and this physical quantity
is known to 15 significant decimal digits!

The states are switched by SFQ volt-time V(t) pulses (or equivalently,
current-inductance I(t) pulses).  Information is transferred with
picosecond pulses generated by the firing of Josephson Junctions,
which, when provided with sufficient bias current, create fluxons (the
quanta of magnetic flux similar to photons for light).  Fluxons are
created by JJs becoming resistive for time sufficient to circumscribe
the flux quantum, and they distribute themselves through
super-conducting circuits obeying the laws of physics.

RSFQ primitive circuit elements must be ready to receive input at any
time.  Our example latch circuit includes JJ $J_3$ which discards
pulses if our latch already contains a fluxon whose current is
circulating through its $J_1$--L--$J_2$ loop.  A pulse on input $C$
will cause $J_3$ to fire and subsequently causes $J_2$ to fire, which
results in the circulating current to escape through the $OUT$
terminal (assuming that a suitable load is connected to its output).  If
a pulse presents itself to input $C$ when the $J_1$--L--$J_2$ loop is
empty, then it will be discarded and no output pulse will appear.

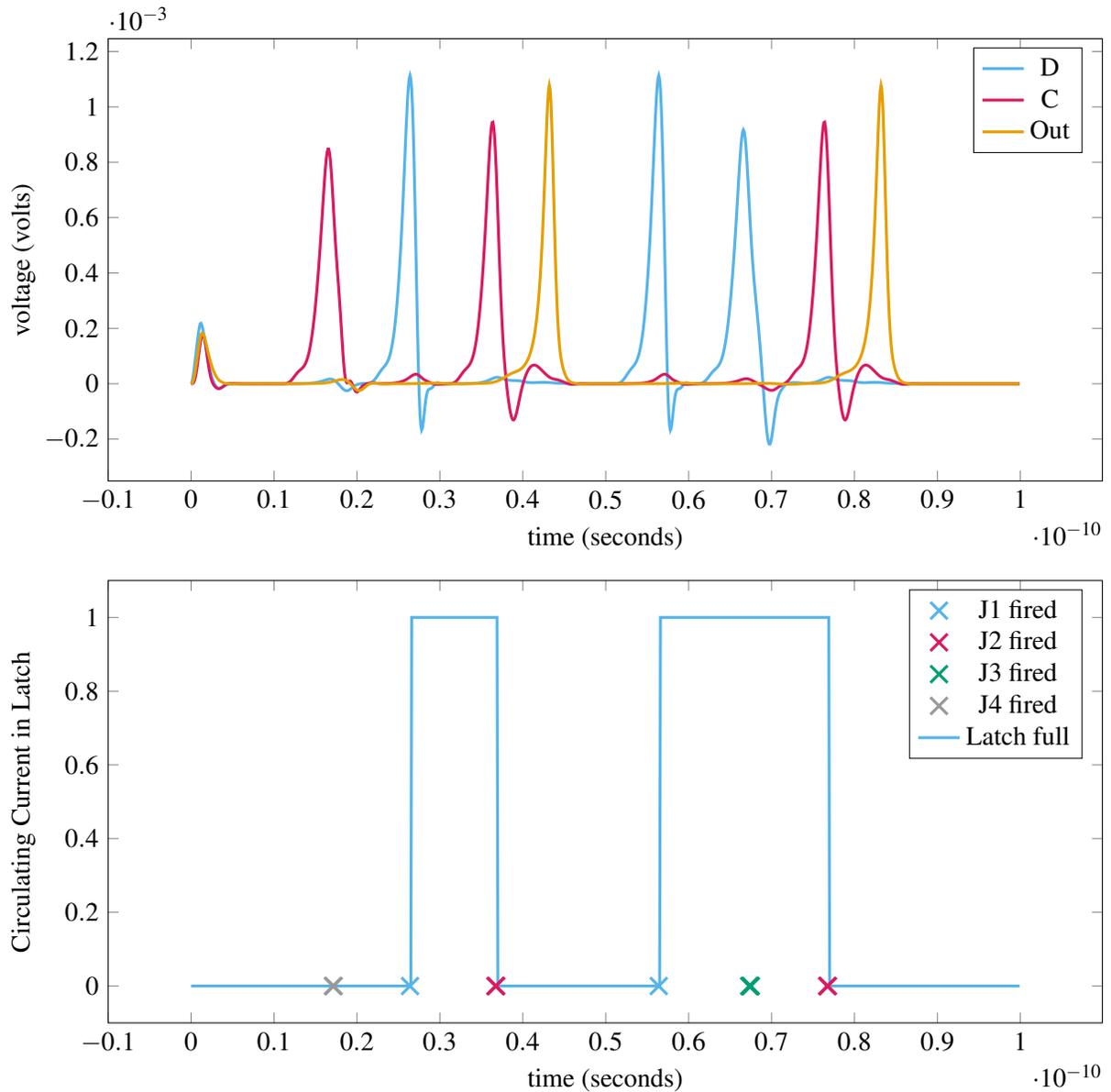
\begin{figure}


\definecolor{magenta}{HTML}{D81B60}
\definecolor{lightorange}{HTML}{E69F00}
\definecolor{lightblue}{HTML}{56B4E9}
\definecolor{pink}{HTML}{CC79A7}
\definecolor{forestgreen}{HTML}{009E73}
\definecolor{grey}{HTML}{999999}

\begin{center}
\begin{tikzpicture}
  \begin{axis}[
      width=16cm,
      height=8cm,
      xlabel = time (seconds),
      ylabel = voltage (volts),
      every axis plot/.append style={very thick}]
    \addplot[color=lightblue] table [x=TIME, y=D, col sep=comma, mark=none] {d-latch-fix.csv};
    \addlegendentry{D}

    \addplot[color=magenta] table [x=TIME, y=C, col sep=comma, mark=none] {d-latch-fix.csv};
    \addlegendentry{C}

    \addplot[color=lightorange] table [x=TIME, y=OUT, col sep=comma, mark=none] {d-latch-fix.csv};
    \addlegendentry{Out}
  \end{axis}
\end{tikzpicture}
\end{center}

\begin{center}
\begin{tikzpicture}
  \begin{axis}[
      width=16cm,
      height=8cm,
      xlabel = time (seconds),
      ylabel = Circulating Current in Latch,
      unbounded coords = jump,
      every axis plot/.append style={very thick}
    ]
    \addplot[only marks, mark=x, mark size=5pt, color=lightblue]
             table [x=$TIME$, y="J1fired", col sep=comma] {timing-diagram-d-latch.csv};
    \addlegendentry{J1 fired}

    \addplot[only marks, mark=x, mark size=5pt, color=magenta]
            table [x=$TIME$, y="J2fired", col sep=comma] {timing-diagram-d-latch.csv};
    \addlegendentry{J2 fired}

    \addplot[only marks, mark=x, mark size=5pt, color=forestgreen]
             table [x=$TIME$, y="J3fired", col sep=comma] {timing-diagram-d-latch.csv};
    \addlegendentry{J3 fired}

    \addplot[only marks, mark=x, mark size=5pt, color=grey]
             table [x=$TIME$, y="J4fired", col sep=comma] {timing-diagram-d-latch.csv};
    \addlegendentry{J4 fired}
    \addplot[color=lightblue] table [x=$TIME$, y="LLtiming", col sep=comma]
             {timing-diagram-d-latch.csv};
    \addlegendentry{Latch full}
  \end{axis}
\end{tikzpicture}
\end{center}

\caption{The first plot is a transient analysis for our simple D
  latch; it begins with a simulation artifact.  The calculation of
  signal derivatives during the first few simulation steps is not
  defined mathematically, and thus the apparent initial pulses are
  simulation {\em noise}.  Note the colors: yellow is the output
  signal, latch input data ($D$) is blue, and the read-and-output {\em
    clock} signal ($C$) is red.  The diagram begins with a {\em
    clock}, and no yellow output ($Out$) signal pulse is produced.
  The first blue input ($D$) fills the storage loop; this is followed
  by a red pulse which causes the latch to emit a yellow pulse that
  also empties the latch.  Reading our latch is destructive, removing
  any saved state if present.  Later we see two blue pulses (with the
  first one appearing at about 55 pico seconds); even though there are
  two consecutive blue ($D$) input pulses, when a read ($C$) signal is
  provided, only one output pulse is produced. The second plot is a
  timing diagram of the same simulation produced from the VWSIM
  output. The blue line depicts when the latch is empty or full. The
  ``X'' markers indicate when each of the JJs fired. The diagram shows
  that J1's blue firing fills the latch when empty, J2's red firing
  drains the latch when full, J3's green firing rejects extra ($D$)
  fill requests, and J4's gray firing rejects extra ($C$) read
  requests.}
\label{fig:d-flip-flop-traces}
\end{figure}

The current through the quantizing inductor, L, provides the clearest
indication that the latch has {\em captured} a fluxon.  Such a current
will continue to circulate in the $J_1$--L--$J_2$ loop until it is
encouraged to escape by input $C$ by causing $J_2$ to fire, which
stops the current passing through $J_2$ to stop and for the
circulating circuit to be diverted toward the output.

The VWSIM simulator accepts a textual form of circuits like the one
shown in Figure~\ref{fig:d-flip-flop}.  When given a valid textual
description, VWSIM processes it into a model that can be simulated,
and then records node voltages or phases and component currents
for later processing or display.  VWSIM provides the capability to
write the user-requested voltage, current, or phase outputs to a
file.  Figure~\ref{fig:d-flip-flop-traces} provides the plot of a
VWSIM simulation of the D latch.

Once the simulation is completed, the VWSIM simulation state and
output are stored persistently in STOBJs available to the interactive
ACL2 loop.  Since the simulation output is stored, users can perform a
wide range of analyses on the simulated circuit.  Consider the
waveforms shown in the first plot of Figure
\ref{fig:d-flip-flop-traces}.  These voltage waveforms demonstrate
what we expect as the input and output of the circuit, but they
conceal whether the devices inside the D latch are working as
intended.  As mentioned earlier, if a D latch is functioning properly
it is expected to have the following properties:
\begin{itemize}
\item
J1 fires when the latch is empty and a D signal is provided, which
causes a circulating current in the $J_1$--L--$J_2$ loop;
\item
J2 fires when the latch is full and a clock signal is provided, which
drains the circulating current;
\item
J3 fires to reject extra pulses when the latch is full; and
\item
J4 fires to reject a C request when the latch is empty.
\end{itemize}
Each of these properties can be confirmed with the help of a timing
diagram. The second plot in Figure~\ref{fig:d-flip-flop-traces} shows
a timing diagram that was produced using VWSIM output.  This timing
diagram indicates that each JJ in the D latch fires correctly to
either empty or fill the latch.

During the development of VWSIM, we have used the simulator to perform
many of these types of analyses.  The results are informing the
decisions we make to define primitives and abstractions that will be
used for the verification of RSFQ circuit models.



\section{Future Work} \label{sec:future-work}


The development of VWSIM may have spawned more ACL2 development work
than the core effort itself.  To achieve useful simulation
performance, we permitted the storage and use of floating-point
numbers.  In addition, our matrix-solve code is based on the ACL2 {\tt
  LOOP\$} construct for which we have little experience; guard proofs
for the loop-with-loop, matrix manipulation code is complex and there
isn't any available library support.

With respect to performance, VWSIM is much slower than JoSIM.  We have
attempted to address this, but some issues are deeper than even the
definition of the ACL2 system -- they have to do with Lisp systems on
which ACL2 is implemented.  It is difficult to arrange for
double-precision (64-bit), floating-point numbers to be used without
them being {\em boxed} by the underlying Lisp systems on which ACL2
operates.  The primary Lisp systems we use are Clozure Common Lisp
(originally named Coral Common Lisp) (CCL~\cite{CCL}) and Steel Bank
Common Lisp (SBCL~\cite{SBCL}).  On 64-bit systems, these two Lisp
systems use part of 64-bit words (4 bits on CCL and 1 bit on SBCL) to
indicate a (partial) data type.  Because double-precision,
floating-point numbers are defined to use 64 bits, additional storage
must be allocated to indicate that a floating-point number is being
referenced.  Providing thorough type declarations helps, but much of
this {\em boxing} activity occurs deep within host Lisp systems and
beyond an ACL2 user's ability to control.

The ACL2 system authors (Kaufmann and Moore) are investigating the
addition of double-precision, floating-point numbers as first-class
ACL2 data objects.  Such floating-point numbers will be forbidden to
appear in a list, but they will be allowed to be stored/retrieved from
STOBJs where such objects have been declared to be floating-point
numbers.  Further declarations will likely be needed when
passing/returning floating-point numbers as arguments/return-values
to/from ACL2 functions.  When using floating-point numbers in {\tt
  LET} and {\tt LET*} expressions, additional declarations will be
required.  Any initial inclusion of floating-point numbers into ACL2
will require careful declaration and will not, initially, include
much, if anything, in the way of semantics.  For instance, negating
the value of a floating-point number may be known to produce the
negative of the original value, but the associativity of addition for
three floating-point values will not be a theorem because there are
many examples where it is not valid.

With respect to performance-oriented operations we can control, we can
improve the performance of symbolic expression evaluator: currently we
have eliminated duplicated evaluation, but we do not perform lazy
evaluation of ``if'' expressions.  We can improve the access speed of
the simulation values we store and retrieve.


We want to develop a means to specify and perform symbolic
verification of RSFQ-based circuits.  We have experience in developing
a hardware-description language that has been used to validate the
FM9001 microprocessor~\cite{FM9001} and has also been used
commercially~\cite{Royal_Society_2017}.  The cost of simulation in the
number of circuit elements grows with the cube of the model size.  Any
such specification and verification system for RSFQ logic will need
some kind of abstract way to specify RSFQ circuit behaviors.
Presently, the VWSIM simulation operates at the level of circuit
components (e.g., inductors, JJs), and the level of detail is too
great to allow simulation of even medium-sized systems.  We have been
thinking about abstractions for RSFQ circuit models like timing
diagrams in Figure \ref{fig:d-flip-flop-traces}.
\section{Conclusion}    

We have defined the VWSIM, circuit simulator.  Our approach involves
solving repeatedly for $x$ in matrix equation $Ax = b$.  When the $A$
matrix does not change, we do not need to re-factor $A$, but we can
use the triangularization of $A$ from an earlier time step.  To
provide support for transcendental functions, we have used an ACL2
{\em trick} to appeal to floating-point functions such as {\em sine}
and {\em exponentiation}; special provisions have been made so we may
use floating-point numbers in ACL2.  VWSIM stops when the simulated
time reaches it end time.  Results are stored for each time step;
VWSIM users are free to further process the results.

We continue to work with the authors of ACL2 to encourage them to
include floating-point numbers as a native ACL2 data type.  This work
is ongoing, and coding the matrix-solve algorithm with the ACL2 {\tt
  LOOP\$} macro as opposed to recursive functions appears to provide
better performance.  We note, however, that the guard conditions for
nested {\tt LOOP\$} statements can be quite complex and libraries
to aid with the definition of functions including {\tt LOOP\$}
references are still under development.

Being able to use ACL2 to script the use of the VWSIM simulator is a
great asset.  For now, we have been using the same stimulus functions
provided by JoSIM, but we are exploring the generation of simulation
stimulus from higher-level specification.  And, instead of graphing
the output of our simulations, we are exploring the use of functions
that recognize valid outputs.  Thus, we are trying to automate the
generation of stimulus and the recognition of valid output.

The development of VWSIM was very helpful in clarifying our
understanding of the modeling and simulation of Josephson Junction
electrical-circuit models.  This understanding led us to find several
bugs in the JoSIM simulation system; having an executable (JoSIM)
specification was a great aid.  Had we developed VWSIM in C, we're
confident that we would still be debugging its operation.  During the
development of VWSIM, we often compared VWSIM results to those
produced by JoSIM.  As we discovered discrepancies, we studied harder
so we could identify what code needed updating; this study improved
our understanding of JJ-based circuits, which will be instrumental in
our work to verify RSFQ circuits models.

\section{Acknowledgments}

This work was supported, in part, by ForrestHunt, Inc. and the
U.S.~Army.  We thank Matt Kaufmann for enabling the use of
floating-point arithmetic in VWSIM, creating VWSIM's SPICE to
S-expression parser, and guiding us on several proofs.  Additionally,
we would like to thank our colleagues Ivan Sutherland, Marly Roncken,
Gary Delp, and Steven Rubin for their support and feedback.

\section{Bibliography}

\bibliographystyle{eptcs}
\bibliography{vwsim-april-2022}

\end{document}